%% file: paper.tex
\documentclass[a4paper,twoside,12pt,notitlepage]{article}

\usepackage[tbtags]{amsmath}
\usepackage{epsf,epsfig,amstext,amssymb,amsthm,latexsym,verbatim,fancyhdr}

\input{header}

\begin{document}

	\input{title}

	\input{abstr}

	\input{introduction}

	\include{chapter2}

	\include{chapter3}

	\include{chapter4}

	\include{conclusion}

\bibliographystyle{unsrt}

\end{document}

%% file: header.tex
\newcommand{\BQN}{\begin{eqnarray}}
\newcommand{\EQN}{\end{eqnarray}}
\newcommand{\BQNY}{\begin{eqnarray*}}
\newcommand{\EQNY}{\end{eqnarray*}}
\newcommand{\BQ}{\begin{equation} }
\newcommand{\EQ}{\end{equation} }
\newcommand{\SU}{{\rm SU}}  
\newcommand{\SO}{{\rm SO}}
\newcommand{\dd}{{\rm d}}  

%% file: title.tex
\title{\textsc{Asymptotics of Relativistic Spin Networks}}
\author{\textsc{John W Barrett} \\ \textsc{Christopher M Steele} \\
\\ School of Mathematical Sciences \\ The University of Nottingham \\
University Park \\ Nottingham NG7 2RD \\ UK}
\date{21 January 2003}

\maketitle

%% file: abstr.tex
\begin{abstract}
The stationary phase technique is used to calculate asymptotic formulae for SO(4) Relativistic Spin Networks. For the tetrahedral spin network this gives the square of the Ponzano-Regge asymptotic formula for the SU(2) 6j-symbol. For the 4-simplex (10j-symbol) the asymptotic formula is compared with numerical calculations of the Spin Network evaluation. Finally we discuss the asymptotics of the SO(3,1) 10j-symbol.
\end{abstract}

%% file: introduction.tex
\section{Introduction}

Spin networks have been used to develop discrete models of quantum gravity
called state sum models. The original state sum model due to Ponzano and
Regge \cite{pr} used tetrahedral spin networks for the group SU(2) (6j-symbols)
glued together to form a three-dimensional space-time manifold. The state
sum model has a partition function which is determined by summing over the
spin labels or parameters in the 6j-symbols, forming a discrete analogue
of the functional integral for quantum gravity. These models were extended
to four-dimensional manifolds by considering spin networks based on a
four-simplex (`10j-symbol'), and simultaneously changing the group to
SO(4), and later, the Lorentz group SO(3,1): the `relativistic' spin
networks \cite{rsn} \cite{lsm}.

A useful tool for analysing the physical content of these models is the
analysis of the asymptotics of the 6j and 10j symbols. This involves
deriving a formula for the behaviour of the values of the spin network
when all the spin labels are simultaneously made large. Ponzano and Regge \cite{pr}
conjectured a beautiful formula for the asymptotics of the 6j symbol
expressed in terms of the geometry of the tetrahedron associated to the
spin labels. This formula is essentially the Einstein-Hilbert action for
the geometric tetrahedon, thus making an explicit connection between the
state sum model and quantum gravity, at least in the asymptotic regime.
The formula was later proved by Roberts \cite{roberts} using the methods of geometric
quantization.

The asymptotic analysis for the 10j symbol was begun in \cite{4simplex}, which
analysed only some of the contributions in a stationary phase
approximation (the non-degenerate ones, see below). Baez, Christensen and Egan \cite{10jalg} \cite{baez-chr:positivity} \cite{baez-chr:spinfoam}
performed some numerical calculations which gave a different scaling
behaviour, indicating that the remaining contributions (the degenerate
ones) not analysed in \cite{4simplex} must be important.

In this paper we consider a fuller asymptotic analysis of the SO(4) 10j
symbol, including the degenerate configurations, to give a complete
picture of the asymptotics, thus explaining the numerical results in \cite{baez-chr:spinfoam}. Some similar results to ours were recently obtained independently
in \cite{baez-chr:10j}.

The paper develops some general machinery for evaluating the asymptotics
of relativistic spin networks. This machinery is applied to two specific
cases: the tetrahedral network, and the 4-simplex (10j symbol). The
importance of considering the tetrahedal network is that the result is
already known, since the relativistic tetrahedral spin network (for SO(4))
is just the square of the SU(2) 6j-symbol. Our asymptotic formula turns out to
be exactly the square of the Ponzano-Regge asymptotic formula. Thus this
gives essentially a simpler proof of the Ponzano-Regge formula than that
given by Roberts. An interesting feature of our analysis, carried out
using the stationary phase approximation to an integral, is that it has
both `degenerate' and `non-degenerate' geometrical configurations. In this
tetrahedral case, these two types of configurations have the same
amplitude and decay at the same rate in the asymptotic region.

Moving on to the SO(4) 10j symbol, we apply a similar analysis but find the
degenerate configurations dominate in the asymptotic region with a slower
decay rate; this explains the numerical results of \cite{baez-chr:spinfoam}.
We analyse the different types of possible stationary points in detail, giving the corresponding decay rates for the asymptotics. Some of the types of stationary point occur for generic values of the spin labels on the 4-simplex while other types are non-generic. In some non-generic cases the degenerate configurations no longer dominate and we identify some of the other configurations which can dominate the asymptotic formula.

The final section discusses the differences between the SO(4) (Euclidean) and the SO(3,1) (Lorentzian) 10j-symbol. The stationary points of the Euclidean case all have analogues in the Lorentzian case; in particular there continue to be non-degenerate stationary points for the Lorentzian 10j-symbol corresponding to geometric 4-simplexes in Minkowski space with spacelike 3-dimensional faces.
However in the generic case the degenerate configurations still dominate the asymptotic formula.

%% file: chapter2.tex
\section{Tetrahedron}

\subsection{Evaluating SO(4) 6j-symbols}

\begin{figure}[h]
\begin{center}
\includegraphics{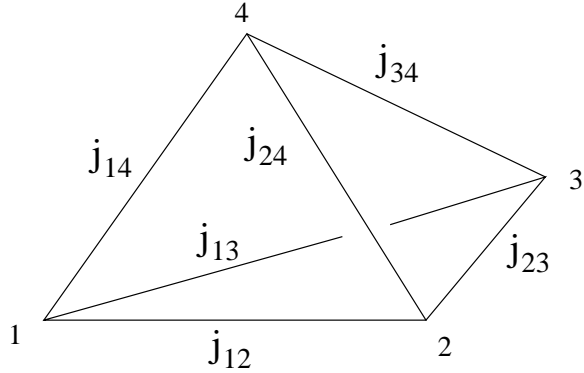}
\end{center}
\caption{Relativistic Spin Network Tetrahedron}\label{fig:tet}
\end{figure}

We begin by calculating the evaluation of the Relativistic Spin Network Tetrahedron. The Relativistic Spin Network is the graph shown in figure~\ref{fig:tet} with edges labelled by half-integer spins. 

The evaluation of this network, a real number, is defined in terms of SU(2) spin networks by 

\begin{equation}
\raisebox{-0.3cm}{$\epsfbox{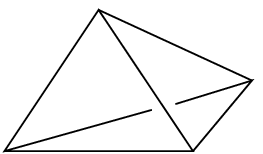}_R$}:= \frac{\epsfbox{smalltet.eps}_{SU(2)} \ \ \ \ \ \ \ \  \epsfbox{smalltet.eps}_{SU(2)}} {\raisebox{-0.8cm}{$\epsfbox{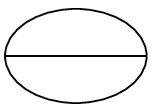}_{SU(2)} \epsfbox{smalltheta.eps}_{SU(2)} \epsfbox{smalltheta.eps}_{SU(2)} \epsfbox{smalltheta.eps}_{SU(2)}$}}.
\end{equation}

In this equation the subscript R indicates a relativistic spin network, and SU(2) indicates an SU(2) spin network. Labels are omitted from the above equation but the edges on both SU(2) tetrahedra have the same labels as the relativistic one on the left hand side. The four theta symbols are formed by connecting each of the four vertices of the tetrahedron to themselves.

The definition of the SU(2) 6j-symbol gives
\begin{equation}
\{6j\}^2_{SU(2)} = \frac{\epsfbox{smalltet.eps}_{SU(2)} \ \ \ \ \ \ \ \  \epsfbox{smalltet.eps}_{SU(2)}} {\left| \raisebox{-0.4cm}{$\epsfbox{smalltheta.eps}_{SU(2)} \epsfbox{smalltheta.eps}_{SU(2)} \epsfbox{smalltheta.eps}_{SU(2)} \epsfbox{smalltheta.eps}_{SU(2)}$} \right|}
\end{equation}
so that the SU(2) 6j-symbol is related to the relativistic spin network evaluation by
\begin{equation}
\raisebox{0.5cm}{$\{6j\}^2_{SU(2)}=(-1)^{\sum_{k<l}2j_{kl}}$} \epsfbox{smalltet.eps}_R.
\end{equation}

Ponzano and Regge gave an asymptotic formula for the 6j-symbol \cite{pr}. Using notation $\{6j\}$ for the value of the 6j-symbol and $PR$ for the Ponzano-Regge approximation we write $\{6j\} \sim PR$ to denote that $\{6j\}$ is asymptotically equal to $PR$ as all the spins are scaled upwards. Ponzano and Regge were not too specific about which scalings of the spins the formula should be valid for. For the formula to be useful it would be necessary to formulate this scaling in the most general possible way that is compatible with proving the result. However this paper is somewhat more preliminary as we are interested in establishing the asymptotic formulae in a restricted set of circumstances to give an outline of the asymptotic behaviour. The scaling that we use is to replace $2j_{kl}+1$ everywhere by $\alpha(2j_{kl}+1)$ for fixed $j_{kl}$, allowing $\alpha\to\infty$ through values where this makes sense (e.g. when $\alpha$ is an integer).

The asymptotic formula means that $\lim_{\alpha \to \infty}(\{6j\}-PR) \alpha^{3/2}=0$. The Ponzano-Regge formula is as follows 
\begin{equation}
PR = \frac{1}{\sqrt{12 \pi V'}}\ \cos\left(\sum_{k<l }\left(j_{kl}+\frac{1}{2}\right) \theta_{kl}+\frac{1}{4}\pi\right)
\end{equation}
where $V'$ is the volume of the tetrahedron with edge lengths $j+\frac{1}{2}$, and $\theta_{kl}$ is the angle between the normal vectors to the faces $k$ and $l$.

Squaring this formula we obtain
\begin{equation}
PR^2 = \frac{1}{3 \pi V}\ \left( 1 + \cos\left(\sum_{k<l}\left(2j_{kl}+1\right) \theta_{kl}+\frac{1}{2}\pi\right)  \right)
\end{equation}
where $V$ is the volume of the tetrahedron with edge lengths $2j+1$.

Notice that $PR^2$ is composed of two parts, a cosine term and a constant term, both of which have the same amplitude. We call the cosine term $(PR^2)_{cos}$ and the constant term $(PR^2)_{const}$ so that
\begin{equation} \label{eq:prsplit}
PR^2=(PR^2)_{const}+(PR^2)_{cos}.
\end{equation}

The evaluation of the relativistic spin network can also be computed using the integral formula \cite{barrett}
\begin{equation} \label{eq:tr}
I=(-1)^{\sum_{k<l} 2j_{kl}} \int_{x \in SU(2)^4} \prod_{k<l} Tr(\rho_{kl}(x_k x_l^{-1})dx
\end{equation}
where $\rho_{kl}$ is the representation of SU(2) on edge $kl$ and $x_k$ is an element of SU(2). We integrate over four $x$'s, one at each vertex of the tetrahedral spin network. The integration uses the Haar measure, normalised to unity. We calculate $I'=I (-1)^{\sum_{k<l} 2j_{kl}}$ so that $I'=\{6j\}^2$.

In the next sections we will rewrite the integral in such a way that the stationary phase formula can be used. Each stationary phase point is a certain configuration for a geometrical tetrahedron (or a generalisation of this). We will show how the different configurations give the two terms in equation (\ref{eq:prsplit}).

\subsection{Kirillov Character Formula}

We begin by rewriting (\ref{eq:tr}) using the Kirillov character formula. 

\begin{equation} \label{eq:kir}
\textrm{Tr}(\rho_{kl}(x_k x_l^{-1}))= \frac{(2j_{kl}+1)r_{kl}}{\sin(r_{kl})} \int_{y_{kl} \in S^2} e^{i (2j_{kl}+1) \xi_{kl}.y_{kl}} \frac{dy}{4 \pi}
\end{equation}
where $\xi_{kl}$ is an element of the Lie algebra of $\SU(2)$ defined by 
\begin{equation} \label{eq:defxi}
\exp(\xi_{kl})=x_l x_k^{-1},
\end{equation}
and $r_{kl}=|\xi_{kl}|$.
The ambiguity in the definition of $\xi$ is fixed by requiring
  $r_{kl}=|\xi_{kl}|$ to be the angle between $x_k$ and $x_l$ thought of as vectors in $S^3\cong\SU(2)$. This gives a unique  $\xi$ for all angles $0\le r<\pi$. The case $r=\pi$, when two $x$'s are anti-parallel, requires special treatment, as the Kirillov formula does not work directly there.  One could use a simple modification of the Kirillov formula which would work in a neighbourhood of $r=\pi$. However consideration of the cases in which $r=\pi$ can be effectively bypassed, as discussed below.

Carrying out the integral in equation (\ref{eq:kir}) gives the Weyl character formula used by Barrett and Williams \cite{4simplex} to find the non-degenerate stationary points.

\begin{equation}
\textrm{Tr}(\rho_{kl}(x_k x_l^{-1}))=\frac{\sin((2j_{kl}+1) r_{kl})}{\sin(r_{kl})}
\end{equation}
It is important here to use the Kirillov formula, rather than the Weyl formula, as it enables us to evaluate the contribution from geometries where $r_{kl}=0$.

\begin{figure}[h]
\begin{center}
\includegraphics{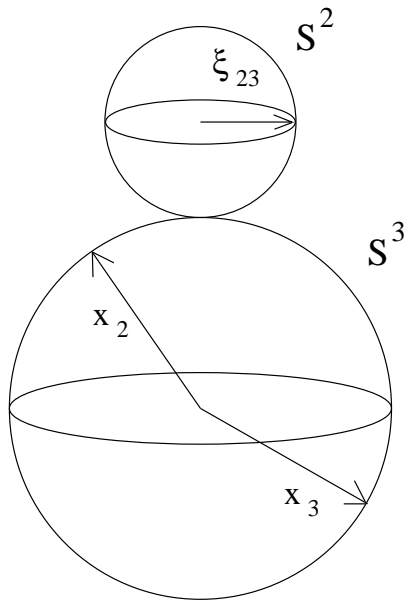}
\end{center}
\caption{Interpretation of the Kirillov Character formula}
\end{figure}

Considering elements $x_{k}$ of SU(2) as vectors in $S^3$, and replacing $2j_{kl}+1$ by $\alpha (2j_{kl}+1)$ to examine the asymptotics when $\alpha \to \infty$, we obtain
\begin{eqnarray} 
I'  & = &  \frac{1}{(4 \pi)^6 (2 \pi^2)^4} \label{eq:tr2} \\
& & \int_{x \in \left(S^3\right)^4} \int_{y \in \left(S^2\right)^6} \left( \prod_{k<l}  \frac{\alpha (2j_{kl}+1) r_{kl}}{\sin(r_{kl})} \right) e^{i \alpha \sum_{k<l}  (2j_{kl}+1) \xi_{kl}.y_{kl}} dy dx \nonumber 
\end{eqnarray}
We call the exponent $\sum_{k<l}  (2j_{kl}+1) \xi_{kl}.y_{kl}$ in equation (\ref{eq:tr2}) the action, and denote it by S.

We consider the integration to be performed in two parts, firstly fixing $x_1$ to be at the north pole of $S^3$, $x_2$ to be in a plane containing $x_1$, and $x_3$ to be in a 3 dimensional hyperplane containing the plane. The second part is to integrate over the symmetry of the integrand (which only depends on relative angles), rotating all the $x$'s and $y$'s at once. 

The symmetry group SO(4) is $6$ dimensional, the original integration was over $24$ dimensions, so the first part of the integral is over $18$ dimensions. Integrating over the symmetries will multiply the result by a volume factor.

\subsection{Method of Stationary Phase}

We use the method of stationary phase \cite{asymptotics} to calculate the asymptotic contributions to the first part of the integral. The general stationary phase formula is
\begin{equation} \label{eq:sp}
\int a(x) e^{i k \phi(x)} dx = \left( \frac{2 \pi}{k} \right)^{n/2} \sum_{x | d\phi(x)=0} a(x) e^{ik\phi(x)} \frac{e^{i \pi \textrm{sgn}(H)/4}}{\sqrt{|\det H(x)|}} + O(k^{-n/2-1})
\end{equation}
where $H(x)$ is the Hessian matrix for $\phi(x)$, and $\textrm{sgn}(H)$ is the number of positive eigenvalues minus the number of negative eigenvalues of $H$.

Writing $I'_{a}$ for the asymptotic approximation to $I'$ i.e. $I' \sim I'_a$, this gives
\begin{equation}
I'_{a}=\frac{2^4 \pi^4}{(4 \pi)^6 (2 \pi^2)^4} \left( \frac{2 \pi}{\alpha} \right)^{18/2} \sum_{x,y | dS=0} \left( \prod_{k<l} \frac{\alpha (2j_{kl}+1)r_{kl}}{\sin(r_{kl})} \right)  \frac{e^{i(\alpha S +\pi \textrm{sgn}(H)/4)}}{\sqrt{|\det H|}}
\end{equation}
where $2^4 \pi^4$ is the volume of the space we have quotiented out of the integral to remove the SO(4) symmetry. The assumption is that in the remaining integral the stationary points are discrete.
Notice that all terms in this formula will scale as $\alpha^{-3}$.

We must find the stationary points of the action
\begin{equation} \label{eq:action}
S=\sum_{k<l} (2j_{kl}+1)\xi_{kl}.y_{kl}
\end{equation} 
with respect to varying $x_{i}$ and $y_{kl}$, remembering that each $\xi_{kl}$ depends on the $x$'s via equation (\ref{eq:defxi}).

There are a number of types of solutions for which the action is stationary.
Firstly the solutions can be classified by the dimension of the linear subspace which is spanned by the unit vectors $x_i$. Secondly, there is a qualitative difference between solutions for which no pair of these vectors are parallel, i.e. $x_i\ne \pm x_j$ for all $i\ne j$, and solutions in which at least one pair are parallel. The former will be called `non-parallel' solutions and the latter `partly parallel' if not all $x_i $ are parallel, and `totally parallel' if $x_i=\pm x_j$ for all $i,j$. Both of these classifications measure the amount of degeneracy of the solutions, and there is some relation between them. In the two examples considered below, the highest possible dimension solutions for a given spin network are necessarily non-parallel. These solutions will be called `non-degenerate'. At the other end of the scale, the totally parallel solutions are clearly the same as the one-dimensional solutions.

For the parallel solutions, one only needs consider the cases when $x_i= x_j$.
Replacing $x_k$  by $-x_k$ has a very simple effect on the value of the integrand in (\ref{eq:tr}): it is multiplied by the factor
$$(-1)^{\sum_{l\ne k} 2j_{kl}}. $$
The factor is equal to $1$ if the parity admissibility condition is satisified, and $-1$ otherwise. This means that if the admissibility condition is not satisfied, then the contributions from stationary points at $x_k$ and $-x_k$ always cancel and so the asymptotic formula always gives zero. If the condition is satisfied then one only need consider one of these two possibilities, and one can always choose $x_i= x_j$ rather than $x_i= -x_j$. Thus one never needs to do a detailed calculation of the contribution of stationary points for which the angle $r=\pi$.

\subsection{Results}
The following sections show that whenever the spins are the edge lengths of a Euclidean tetrahedron
the stationary phase formula has discrete stationary points which are one of two types, either non-degenerate, or one-dimensional. Accordingly, the stationary phase formula can be written
$$ I'_a=C_{non-deg}+C_{1-d}.$$
The contribution from non-degenerate configurations, $C_{non-deg}$, 
gives exactly the oscillatory part of the Ponzano-Regge squared formula 
\begin{equation}
C_{non-deg}=(PR^2)_{cos}.
\end{equation}
The contribution from one-dimensional configurations, $C_{1-d}$, gives exactly the constant part of the Ponzano-Regge squared formula 
\begin{equation}
C_{1-d}=(PR^2)_{const}.
\end{equation}
Putting these two results together, this shows that the 
  asymptotic formula for the relativistic spin network tetrahedron   agrees exactly with the square of the Ponzano-Regge formula whenever we have a tetrahedron which can be embedded into 3-dimensional Euclidean space, 
\begin{equation}
I'_{a}=PR^2.
\end{equation}

\subsection{Non-parallel configurations}

Varying the action with respect to the $y$'s we find that the action is stationary when $y_{kl}$ is parallel or antiparallel to $\xi_{kl}$. In other words,
\begin{equation}
y_{kl}=\epsilon_{kl} \frac{\xi_{kl}}{|\xi_{kl}|},
\end{equation}
with $\epsilon_{kl}=\pm1$.

Varying the action with respect to the $x$'s, and using a Lagrange multiplier $\lambda_i$ for the constraint $x_i.x_i=1$ we obtain four equations, one for each face of a tetrahedron.
\begin{equation}
\begin{aligned}
\frac{\epsilon_{12}(2j_{12}+1)x_{2}}{\sin(r_{12})}+ \frac{\epsilon_{13}(2j_{13}+1)x_{3}}{\sin(r_{13})}+ \frac{\epsilon_{14}(2j_{14}+1)x_{4}}{\sin(r_{14})}
& = & \lambda_1 x_1 \\
\frac{\epsilon_{12}(2j_{12}+1)x_{1}}{\sin(r_{12})}+ \frac{\epsilon_{23}(2j_{23}+1)x_{3}}{\sin(r_{23})}+ \frac{\epsilon_{24}(2j_{24}+1)x_{4}}{\sin(r_{24})}
& = & \lambda_2 x_2 \\
\frac{\epsilon_{13}(2j_{13}+1)x_{1}}{\sin(r_{13})}+ \frac{\epsilon_{23}(2j_{23}+1)x_{2}}{\sin(r_{23})}+ \frac{\epsilon_{34}(2j_{34}+1)x_{4}}{\sin(r_{34})}
& = & \lambda_3 x_3 \\
\frac{\epsilon_{14}(2j_{14}+1)x_{1}}{\sin(r_{14})}+ \frac{\epsilon_{24}(2j_{24}+1)x_{2}}{\sin(r_{24})}+ \frac{\epsilon_{34}(2j_{34}+1)x_{3}}{\sin(r_{34})}
& = & \lambda_4 x_4 
\end{aligned}
\end{equation}

The equations show that there is at least one linear relation between the $x$'s and so
require the $x$'s to lie in a 3-dimensional hyperplane. Taking $v_{kl}$ to be the unit vector in the direction $x_k \times x_l$ (using the vector cross product in the 3d hyperplane) gives
\begin{equation}
\label{eq:nonpar} 
\begin{aligned} 
\epsilon_{12}(2j_{12}+1)v_{12}+ \epsilon_{13}(2j_{13}+1)v_{13}+ \epsilon_{14}(2j_{14}+1)v_{14}
& = & 0  \\
-\epsilon_{12}(2j_{12}+1)v_{12}+ \epsilon_{23}(2j_{23}+1)v_{23}+ \epsilon_{24}(2j_{24}+1)v_{24}
& = & 0 \\
-\epsilon_{13}(2j_{13}+1)v_{13}- \epsilon_{23}(2j_{23}+1)v_{23}+ \epsilon_{34}(2j_{34}+1)v_{34}
& = & 0  \\
-\epsilon_{14}(2j_{14}+1)v_{14}- \epsilon_{24}(2j_{24}+1)v_{24}- \epsilon_{34}(2j_{34}+1)v_{34}
& = & 0  
\end{aligned}
\end{equation}

\subsubsection{Three-dimensional configurations}
\begin{figure}[h]
\begin{center}
\includegraphics{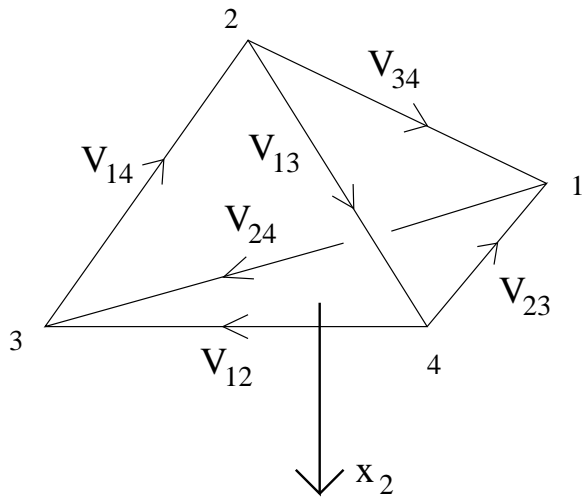}
\end{center}
\caption{Non-degenerate stationary point} \label{fig:ndtet}
\end{figure}

Now consider the case of a three-dimensional solution.
Setting $V_{kl}=\epsilon_{kl}(2j_{kl}+1)v_{kl}$, the above equations show that the $V_{kl}$ are the edge vectors of a tetrahedron with edge lengths $2j_{kl}+1$. Each equation describes the face of a geometrical tetrahedron as shown in figure \ref{fig:ndtet}. Notice that the four equations are not linearly independent - the sum of the first three gives the fourth equation.

Each $x_i$ is a normal vector to a face of the tetrahedron; it could be inward or outward pointing. Let $n_i$ be the outward normal to the $i$-th face. Then $x_i=a_i n_i$ with $a_i=\pm1$.

In figure~\ref{fig:ndtet} the edge vectors for the tetrahedron are given by the expression $\pm(2j_{kl}+1)n_k\times n_l=\pm(2j_{kl}+1) a_k a_l v_{kl}$, with $\pm$ indicating one overall choice of sign. This means that, for figure~\ref{fig:ndtet}, the $\epsilon$ are determined by $\epsilon_{kl}=\lambda a_k a_l$, with $\lambda=\pm1$. This means that many choices of $\epsilon_{kl}$ give no solutions at all.
 
Every stationary point is of this form. Consider any set of $x$'s and $\epsilon$'s satisfying equations (\ref{eq:nonpar}), then this corresponds to a geometric tetrahedron. By acting on the $x$ vectors with an element of SO(4) this tetrahedron can be rotated to correspond to the tetrahedron drawn in figure \ref{fig:ndtet}, possibly with all of the arrows indicating the directions of the $V$ reversed. 

Given one stationary point $\{x_i,\epsilon_{kl}\}$ we can generate all of the others using combinations of two operations. The first operation is to swap the sign of one or more of the x's. Since $x_1$ is fixed at the north pole we are not at liberty to swap its sign. For $\sigma_i=\pm 1$,

\begin{eqnarray*}
x_1 & \to & x_1 \\
x_2 & \to & \sigma_2 x_2 \\
x_3 & \to & \sigma_3 x_3 \\
x_4 & \to & \sigma_4 x_4 \\
\epsilon_{kl} & \to & \sigma_k \sigma_l \epsilon_{kl}.
\end{eqnarray*}

The second operation is to simultaneously swap the sign of all of the $\epsilon$ whilst leaving the $x$ unchanged

\begin{eqnarray*}
x_i & \to & x_i \\
\epsilon_{kl} & \to & - \epsilon_{kl}
\end{eqnarray*}

If $\sum_{vertex} j_{kl}$ is an integer, each of the 8 stationary points related by the first operation gives the same contribution to the integral.  The stationary points which are related to these by the second operation give the complex conjugate contribution. 

If $\sum_{vertex} j_{kl}$ is not an integer (i.e. is an odd half-integer), the stationary points for the $x$'s cancel and there is no contribution from the non-degenerate stationary points. This corresponds to the admissibility condition for spin networks. For a spin network to be called admissible the $j$'s labelling the edges incident to each vertex must sum to an integer (and satisfy triangle inequalities). The admissible spin networks are the only ones with non-zero evaluation. If the admissibility condition is not met there will be no stationary point contribution. The stationary phase calculations also give the triangle inequalities for spin networks, since to be able to form the triangular faces of a tetrahedron each spin in a face must not be greater than the sum of the other two.

The Hessian is an $18 \times 18$ matrix with components $H_{kl}=\frac{\partial^2 S}{\partial u_k \partial u_l}$ where $u_1 \ldots u_{18}$ are the coordinates we are integrating over. For any particular 6j-symbol the eigenvalues of the Hessian can be found using computer algebra.  The Hessian in the $\epsilon$'s all positive case has 2 positive eigenvalues and 16 negative eigenvalues. Hence $\textrm{sgn}(H)=-14$. The $\epsilon$'s all negative case has $\textrm{sgn}(H)=14$.

These contributions will combine to give an asymptotic expression of the form

\begin{eqnarray}
C_{non-deg}=\frac{A}{2} \left( e^{i(\sum_{k<l}(2j_{kl}+1)r_{kl}-14\pi/4)}+
e^{i(-\sum_{k<l}(2j_{kl}+1)r_{kl}+14\pi/4)}\right) \nonumber \\=A \cos\left(\sum_{k<l}(2j_{kl}+1)r_{kl}+\pi/2\right)
\end{eqnarray}

The determinant of the Hessian can be evaluated using computer algebra for any particular 6j-symbol. The amplitude of the contribution has been calculated for a large number of cases and compared with the amplitude of the cosine term in the square of the Ponzano Regge formula. Examining a large number of cases, we find that they are always equal:
\begin{equation}
A=\frac{1}{3 \pi V}.
\end{equation}
More precisely, the amplitudes agree exactly for all Euclidean tetrahedra, which are refered to as type I in Ponzano-Regge \cite{pr}. Ponzano and Regge's type II and III tetrahedra, the Lorentzian and transition cases do not have an analogue here.  However we have been unable to prove this formula in general, as the algebraic expressions for the determinant of the Hessian are too large. 

The final result is
\begin{equation}
C_{non-deg}=\frac{1}{3 \pi V} \cos\left(\sum_{k<l}(2j_{kl}+1)r_{kl}+\pi/2\right).
\end{equation}

\subsubsection{Non-parallel 2-dimensional configurations}

The only lower dimensional non-parallel configurations possible are where all the $x$'s lie in a 2 dimensional hyperplane. If the admissibility conditions for the tetrahedral spin network are satisfied then there are no 2 dimensional solutions to the stationary phase equations (\ref{eq:nonpar}). 

If the admissibility conditions are not satisfied then the contributions for different $\epsilon$'s exactly cancel.
Hence the lower dimensional configurations do not contribute to the asymptotic formula.

\subsection{One-dimensional configurations}

The one-dimensional configurations involves all $x$'s parallel or antiparallel.
As noted above, one need only consider the case when all $x$'s are actually equal, the other cases giving the same asymptotics.

\begin{figure}[h]
\begin{center}
\includegraphics{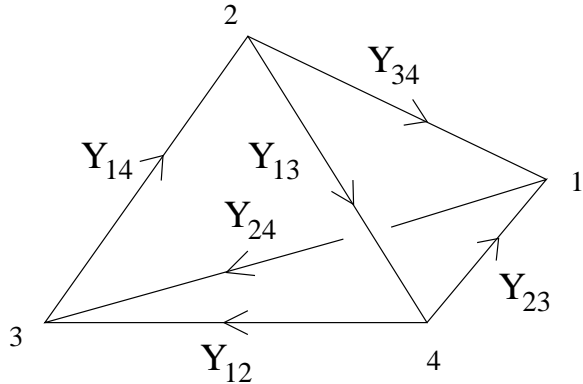}
\end{center}
\caption{Degenerate stationary point} \label{fig:dtet}
\end{figure}

Returning to the action (\ref{eq:action}) if we vary the $y$'s the action is stationary in all cases. If we vary the $x$'s we find that for the action to be stationary
\begin{equation}
\begin{aligned}
(2j_{12}+1)y_{12}+(2j_{13}+1)y_{13}+(2j_{14}+1)y_{14}
& = & 0 \\
-(2j_{12}+1)y_{12}+(2j_{23}+1)y_{23}+(2j_{24}+1)y_{24}
& = & 0 \\
-(2j_{13}+1)y_{13}-(2j_{23}+1)y_{23}+(2j_{34}+1)y_{34}
& = & 0 \\
-(2j_{14}+1)y_{14}-(2j_{24}+1)y_{24}-(2j_{34}+1)y_{34}
& = & 0
\end{aligned}
\end{equation}
This means the $y$'s are parallel to the edge vectors of the same tetrahedron we found in the non-degenerate case. Indeed, setting $Y_{kl}=(2j_{kl}+1)y_{kl}$ we now have the $Y_{kl}$ as edge vectors of a tetrahedron with edge lengths $2j+1$, as in figure \ref{fig:dtet} (or with the direction of the arrows swapped)

There will only be a solution to the stationary phase equations if the j's satisfy the triangle inequalities. There are $2^3$ stationary points of this type for the $x$'s (since we can choose each of $x_2 \ldots x_4$ to be parallel or antiparallel to $x_1$) and 2 sets of $y$'s for each (the vectors could be as drawn in figure \ref{fig:dtet} or reversed). The contributions are identical if the parity admissibility condition is satisfied, and exactly cancel if not.

The Hessian for the all $x$'s parallel case has 9 positive and 9 negative eigenvalues. Hence $\textrm{sgn}(H)=0$. These contributions will be non-oscillating, since $S=0$ for one-dimensional configurations.

The amplitude of the contribution has been calculated for a number of cases, and compared with the amplitude of the constant term in the square of the Ponzano Regge formula. Again we find that in all cases
\begin{equation}
A=\frac{1}{3 \pi V}.
\end{equation}

Denoting the contribution from one-dimensional configurations by $C_{1-d}$

\begin{equation}
C_{1-d}=\frac{1}{3 \pi V}
\end{equation}

\subsection{Higher-dimensional parallel configurations}

There is never any contribution from partly parallel configurations where some but not all of the $x$'s are parallel. Again if the admissibility conditions for the tetrahedral spin network are satisfied then there are no solutions to the stationary phase equations. If the admissibility conditions are not satisfied then the contributions for different $\epsilon$'s exactly cancel.

%% file: chapter3.tex
\section{4-Simplex}

\subsection{Evaluating SO(4) 10j-symbols}

We now repeat the previous calculation for the Relativistic Spin Network 4-simplex, or SO(4) 10j-symbol.  

\begin{figure}[h]
\begin{center}
\includegraphics{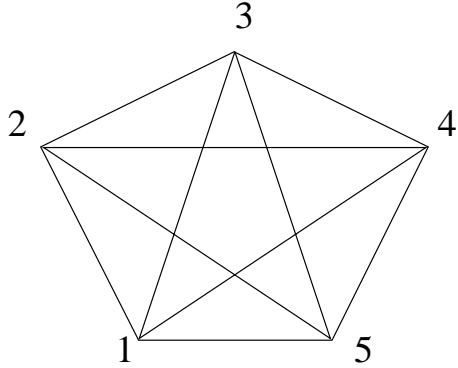}
\end{center}
\caption{Relativistic Spin Network 4 Simplex}
\end{figure}

We use the same integral formula, now with 5 vertices and 10 edges.

\begin{equation} \label{eq:trs}
I=(-1)^{\sum_{k<l} 2j_{kl}} \int_{x \in SU(2)^5} \prod_{k<l} Tr(\rho_{kl}(x_k x_l^{-1})dx
\end{equation}

Set $I'=(-1)^{\sum_{k<l}2j_{kl}}I$ to simplify the equation.
Replacing $2j_{kl}+1$ by $\alpha (2j_{kl}+1)$ to investigate the asymptotics, and using the Kirillov character formula we obtain

\begin{eqnarray} 
I' & = & \frac{1}{(4 \pi)^{10} (2 \pi^2)^5} \label{eq:trs2} \\
& & \int_{x \in \left(S^3\right)^5} \int_{y \in \left(S^2\right)^{10}} \left( \prod_{k<l} \frac{\alpha (2j_{kl}+1)r_{kl}}{\sin(r_{kl})} \right) e^{i \alpha \sum_{k<l} (2j_{kl}+1)\xi_{kl}\cdot y_{kl}} dy dx \nonumber
\end{eqnarray}

For the 10j symbol many of the stationary points are no longer points but higher dimensional manifolds. We integrate over this space of stationary points as for the symmetry group in chapter 2. Each degree of freedom (i.e. dimension) in the stationary phase solution manifold reduces the size of the Hessian by 1, and hence changes the scaling behaviour of the asymptotic contribution by $\alpha^{\frac{1}{2}}$. We are integrating over 35 dimensions (15 for the $x$ variables and 20 for the $y$ variables), and there are always 6 symmetry degrees of freedom from $SO(4)$. If there are an additional $n$ degrees of freedom in the stationary phase solution then the amplitude of the contribution will be
\begin{equation}
\alpha^{-\frac{35-6-n}{2}}\alpha^{10}=\alpha^{-\frac{9-n}{2}}
\end{equation}
where the factor of $\alpha^{10}$ comes from the product in the integral.

The stationary points are the stationary points of the action

\begin{equation} \label{eq:actions}
S=\sum_{k<l} (2j_{kl}+1)\xi_{kl}\cdot y_{kl}.
\end{equation} 

We consider the contribution from non-de\-gen\-er\-ate, lower dimensional, and parallel configurations.

\subsection{Results}
The following sections show that the stationary points in the asymptotic formula are one of a number of types. For a generic set of spins the scaling behaviour of these types are as follows. The one-dimensional stationary points give a contribution which scales as $\alpha^{-2}$. The four- and three-dimensional non-parallel stationary points give contributions which scale as $\alpha^{-9/2}$, although it is possible that in some cases one or both of these types may be absent (due to inequalities which the spins must satisfy). The partly parallel stationary points have only one generic contribution (again subject to inequalities). This is a three-dimensional stationary point with just two of the $x$ variables parallel. Again this scales as $\alpha^{-9/2}$.

This means that that asymptotically, the evaluation of the 10j symbol is dominated by the degenerate one-dimensional stationary points, whose contribution in general decays as a constant times $\alpha^{-2}$, 
\begin{equation}I'_a \sim K\alpha^{-2}.
\end{equation}
 
This is in contrast to the 6j symbol case where the contribution from the degenerate and non-degenerate stationary points had equal amplitude.

The are some non-generic cases of spins where this behaviour can differ, as mentioned in the text below. For example, at the limit of admissibility, the scaling of the one-dimensional configurations can increase to $\alpha^{-7/2}$ or this type may be absent; in these cases it is expected that the dominant contribution will come from a non-generic partly-parallel configuration. For one or two vertices at the limit of admissibility, these configurations give an asymptotic contribution which scales as $\alpha^{-3}$. This agrees with the numerical results obtained by Baez, Christensen and Egan\cite{baez-chr:spinfoam}. Other examples of non-generic behaviour are detailed in the following sections.

\subsection{Non-degenerate 4 dimensional configurations}

First we consider non-degenerate solutions where the $x_{i}$ have a 4 dimensional span, following \cite{4simplex}.

Varying the action with respect to the $y$'s we find that the action is stationary when $y_{kl}$ is parallel or antiparallel to $\xi_{kl}$. Let $\epsilon_{kl}=\pm1$ then

\begin{equation}
y_{kl}=\epsilon_{kl} \frac{\xi_{kl}}{|\xi_{kl}|}
\end{equation}
The action at the stationary points in the $y$'s becomes
\begin{equation}
S=\sum_{k<l}\epsilon_{kl}(2j_{kl}+1)r_{kl}
\end{equation}
A set of 5 non-degenerate vectors with a 4 dimensional span determine a geometric 4-simplex up to scale \cite{rsn}. For a geometric 4-simplex we have Schl\"afli's identity
\begin{equation}
\sum_{k<l}A_{kl} \, d\phi_{kl}=0
\label{eq:schlafli}
\end{equation}
where $\phi_{kl}$ are the angles between the outward normal vectors $n_k,n_l$ to the tetrahedra $k$ and $l$ of the 4 simplex. The $A_{kl}$ are the areas of the triangles of a geometric 4-simplex determined by the $\phi_{kl}$ (up to overall scaling of the 4 simplex). Varying the action, using a Lagrange multiplier $\lambda$ for the constraint equation (\ref{eq:schlafli}), gives
\begin{equation}
dS=\sum_{k<l}\epsilon_{kl}(2j_{kl}+1) \, dr_{kl}=\lambda\sum_{k<l}A_{kl} \, d\phi_{kl}
\end{equation}

If we take the $x$'s to be the outward normals to the tetrahedra of the 4 simplex then $r_{kl}=\phi_{kl}$. If an $x$ is swapped to be an inward normal then $r_{kl}=\pi-\phi_{kl}$; hence $dr_{kl}=-d\phi_{kl}$ for all $r$'s involving the inward $x$.

Taking $n_{k}$ to be the outward normal, and introducing variables $a_{k}$ which are $+1$ if $x_k$ is outward and $-1$ if $x_k$ is inward pointing, we have $x_k=a_k n_k$. For each triangle we obtain

\begin{equation}
\epsilon_{kl}(2j_{kl}+1)=\lambda a_k a_l A_{kl}
\end{equation}

Taking $\lambda=\pm 1$ to fix the overall scale of the $A$'s we see that $A_{kl}=(2j_{kl}+1)$ and $\epsilon_{kl}=\lambda a_k a_l$ for each $k,l$. Hence at a non-degenerate stationary point the $r_{kl}$ are the angles between the inward or outward pointing normals to a 4 simplex with areas $(2j+1)$.

If the parity admissibility conditions (sum of $j$'s at every vertex is an integer) are satisfied for the 10j-symbol, each solution for $\lambda=+1$ gives the same contribution, and each $\lambda=-1$ solution gives the complex conjugate contribution. These terms will combine to give a cosine contribution to the integral. There will be $2^4$ distinct stationary points for the $x$'s ($\frac{2^5}{2}$ because there are 5 $a$'s which can each be $\pm1$ but swapping the sign of them all corresponds to $-I \in \textrm{SO(4)}$) and 2 possible sets of $\epsilon$'s for each, coming from the choice of $\lambda$. Changing  $\lambda$ changes the sign of all of the $\epsilon$'s, swapping the direction of all of the $y_{kl}$.

If the parity admissibility conditions are not satisfied the terms will cancel giving no net contribution. Inequalities analogous to the triangle inequalities for vertices in the 6j-symbol will occur in this case because to be able to form a tetrahedron with four triangles of area $(2j+1)$, each area must not be greater than the sum of the other three.

There may in general be more that one geometric 4 simplex with a given set of areas, for example see Tuckey's example in \cite{barrett:regge} which is a Euclidean 4-simplex for $t<8/3$. When this occurs each possible configuration will have a contribution to the asymptotic formula. From one solution to the stationary phase equations swapping $x_4$ and $x_5$ will yield another solution \cite{baez-chr:10j}. In general these are the only pair of vectors which can be interchanged like this since $x_1 \ldots x_3$ are constrained to lie in hyperplanes.

Terms from the Hessian and the integrand show that the amplitude of the contribution from the non-degenerate stationary points will scale as $\alpha^{-\frac{9}{2}}$.

\subsubsection{Example: Regular 4-simplex}

For the regular 4 simplex with $\alpha(2j_{kl}+1)=\beta$ for all $k,l$ we get a contribution $C_{non-deg}$ from non-degenerate geometric 4-simplexes. A corollary of Bang's theorem \cite{bang} tells us that there is only one geometric 4 simplex with all areas equal \cite{baez-chr:10j}. The corresponding contribution has been calculated by computing the value of the non-exponential part of the integrand and the Hessian at the stationary points.

\begin{equation}
C_{non-deg}= \left( \frac{(24)  3^{\frac{3}{4}}5^{\frac{1}{4}}  \beta^{-\frac{9}{2}}}{25 \pi^\frac{3}{2}} \right) \cos\left(10 \beta \cos^{-1}(-\frac{1}{4})+\frac{\pi}{4}\right)
\end{equation}

\subsection{One-dimensional configurations}

For this case we take all the $x$'s to be parallel or antiparallel. Each of the $2^4$ cases has an identical contribution to the integral. The action is always stationary with respect to varying the $y$'s in this case. Varying the action with respect to the $x$'s gives five equations in the $y$ variables.

\begin{equation}
\label{eq:4sd} 
\begin{aligned}
(2j_{12}+1)y_{12}+(2j_{13}+1)y_{13}+(2j_{14}+1)y_{14}+(2j_{15}+1)y_{15}
& = & 0 \\
-(2j_{12}+1)y_{12}+(2j_{23}+1)y_{23}+(2j_{24}+1)y_{24}+(2j_{25}+1)y_{25}
& = & 0 \\
-(2j_{13}+1)y_{13}-(2j_{23}+1)y_{23}+(2j_{34}+1)y_{34}+(2j_{35}+1)y_{35}
& = & 0 \\ 
-(2j_{14}+1)y_{14}-(2j_{24}+1)y_{24}-(2j_{34}+1)y_{34}+(2j_{45}+1)y_{45}
& = & 0 \\
-(2j_{15}+1)y_{15}-(2j_{25}+1)y_{25}-(2j_{35}+1)y_{35}-(2j_{45}+1)y_{45}
& = & 0 
\end{aligned}
\end{equation}

\begin{figure}[h]
\begin{center}
\includegraphics{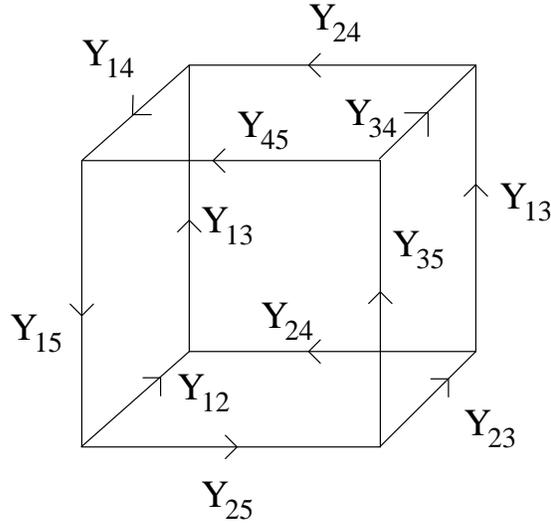}
\end{center}
\caption{4-simplex totally parallel stationary phase point} \label{fig:cube}
\end{figure}

Setting $Y_{kl}=(2j_{kl}+1)y_{kl}$ these equations correspond to the vector diagram in figure \ref{fig:cube}. The front, top, bottom, left and right faces each correspond to a stationary phase equation above. The back face is added in to complete the figure.

Generally the $y$ vectors subject to the above set of equations will have 8 remaining degrees of freedom. This is because there are 20 $y$ variables and 12 independent equations. These 8 degrees of freedom consist of 3 from rotations in SO(3) and 5 deformations of the above `cube'. The angles in the cube in figure \ref{fig:cube} can be varied as long as the edge lengths remain unchanged.

To see that there are 5 degrees of freedom in deforming the `cube' notice that there is one angle to determine the geometry of the back parallelogram face, then 2 spherical coordinate angles each will determine the directions of two diagonally opposite vectors between the back face and the front face ($y_{14}$ and $y_{23}$ say). Once these choices have been made the rest of the vectors are fixed up to a finite set of discrete possibilities, by requiring the edge lengths to be $2j_{kl}+1$.  There will be certain inequalities which must be satisfied to ensure that the figure can be embedded in Euclidean space. Thus the parallel stationary point contribution is of the form $\alpha^{-\frac{9-5}{2}}=\alpha^{-2}$.

For a stationary phase solution of this type to exist and for the contributions from parallel and antiparallel terms to add up requires satisfaction of the admissibility conditions for the 10j-symbol.

These results agree with numerical calculations of 10j symbols carried out by Baez, Christensen and Egan. In \cite{10jalg} and \cite{baez-chr:spinfoam} numerical values of the 10j symbols are calculated and discussed. They see 10j symbols decaying as $O(\alpha^{-2})$. 

\subsection{Partly parallel configurations}

We must consider the case where some but not all of the $x$'s are parallel. There are a number of ways this may happen as there can be one or two directions each parallel to two or more vectors. Unlike the 6j-symbol, for the 10j-symbol contributions may come from configurations of these types, depending on the labelling by spins. 

The only generic case is when just two of the $x$ vectors are parallel. Then the $x$ span 3 dimensions and are the normal vectors to a tetrahedron. For this stationary point to exist inequalities between the spins must be satisfied so that the tetrahedron may be embedded in 3 dimensional Euclidean space. The $x$ and $y$ vectors are uniquely determined by the stationary phase equations so there are no degrees of freedom. The contribution from this configuration will scale as $\alpha^{-\frac{9}{2}}$.

Non-generically, with two vectors parallel, if the spins satisfy a set of particular equalities, it is possible for one of the stationary phase equations to become trivial, allowing up to 4 degrees of freedom.
For $x_1$ and $x_2$ parallel this non-generic stationary point occurs only if
\begin{equation}\label{eq:pair}
\begin{aligned}
j_{13}&=&j_{23}\\
j_{14}&=&j_{24}\\
j_{15}&=&j_{25}\\
j_{34}&=&j_{35}&=&j_{45}.
\end{aligned}
\end{equation}
 This gives a contribution scaling as $\alpha^{-\frac{5}{2}}$.

If two pairs of $x$ vectors are separately parallel there can be a stationary point only if the spins satisfy certain equalities.
If $x_1$ and $x_2$ are parallel, and $x_3$ and $x_4$ are separately parallel then this stationary point occurs only if there exists a set of $\epsilon$ such that the $j$'s satisfy
\begin{eqnarray}
&j_{15}=j_{25}& \nonumber\\
&j_{35}=j_{45}&\\
&\epsilon_{13}(2j_{13}+1)+\epsilon_{14}(2j_{14}+1)+\epsilon_{23}(2j_{23}+1)+\epsilon_{24}(2j_{24}+1)=0. \nonumber
\end{eqnarray}
The $x$ vectors span 3 dimensions and have up to 3 degrees of freedom, whilst the $y$'s are fixed. The contribution from this configuration, when it occurs, will scale as $\alpha^{-3}$.

Three $x$ vectors parallel can be a stationary point only as a special case of the two-dimensional configuration. The spins must satisfy a number of equalities so this is not a generic contribution. If there is no 2 dimensional stationary point with the $x$'s non-parallel then the contribution from this configuaration when it exists will be of order $\alpha^{-\frac{7}{2}}$. If there is a 2 dimensional non-parallel stationary point then this configuration will just be part of that stationary point. The $y$'s are uniquely determined by the stationary phase equations.  The same happens if three $x$ vectors are parallel and the remaining pair of $x$ vectors are separately parallel, however the contribution when there is no 2 dimensional stationary point will be of order $\alpha^{-4}$.

Finally, four of the $x$ vectors may be parallel. This is a non-generic stationary point since there is an equality between the spins. 
If $x_1,x_2,x_3,x_4$ are parallel then this stationary point occurs only if there exist a set of $\epsilon$ such that
\begin{equation}\label{eq:fourp}
\epsilon_{15}(2j_{15}+1)+\epsilon_{25}(2j_{25}+1)+\epsilon_{35}(2j_{35}+1)+\epsilon_{45}(2j_{45}+1)=0.
\end{equation}
There are up to 2 degrees of freedom in the $y$ variables and one degree of freedom in the $x$ variables, so the contribution from this stationary point will generally scale as $\alpha^{-3}$.

\subsection{Non-parallel lower dimensional configurations}

There may also be stationary phase points for configurations in which the $x_{i}$ are not parallel but lie in a hyperplane of dimension less than 4.

\subsubsection{3 dimensional configurations}

Consider configurations where $x_i$ lie in a 3 dimensional hyperplane but no pairs are parallel. 

Using the vector cross product in the 3 dimensional hyperplane we write $v_{kl}$ for the unit vector in the direction $x_k \times x_l$, and $V_{kl}=\epsilon_{kl}(2j_{kl}+1)v_{kl}$. The stationary phase equations show that $y_{kl}=\epsilon_{kl} \xi_{kl}$ and
\begin{equation}
\label{eq:4s3d} 
\begin{aligned}
V_{12}+V_{13}+V_{14}+V_{15}
& = & 0 \\
-V_{12}+V_{23}+V_{24}+V_{25}
& = & 0 \\
-V_{13}-V_{23}+V_{34}+V_{35}
& = & 0 \\ 
-V_{14}-V_{24}-V_{34}+V_{45}
& = & 0 \\
-V_{15}-V_{25}-V_{35}-V_{45}
& = & 0
\end{aligned}
\end{equation}

Notice that these equations are very similar to those obtained for the totally parallel case in (\ref{eq:4sd}). However since $v_{kl}$ is orthogonal to $x_k$ for each $l$ we see that each equation in (\ref{eq:4s3d}) relates vectors which lie in a plane. The geometric figure differs from the totally parallel case because here all its faces are planar. The geometry is more constrained than in the totally parallel case.

The geometric interpretation of a solution to these equations is a degenerate case of a 4 simplex in which every triangle shares a common direction. 

The figure in the totally parallel case had 5 degrees of freedom. Now we are looking at the same figure with the additional constraint that each of 5 faces must be planar. Therefore we expect that when solutions exist to these equations there will be no further degrees of freedom available, and that the contribution to the integral will be $\alpha^{-\frac{9}{2}}$ multiplied by some oscillating function of modulus 1.

\subsubsection{2 dimensional configurations}

Consider configurations where $x_i$ are not parallel but lie in a 2 dimensional hyperplane. The unit bivectors formed from the $x_i$ will be equal (up to sign). Write $\delta_{kl}$ for the sign of $x_k \wedge x_l$ relative to $x_1 \wedge x_2$ ($\delta_{12}=1$).

Then the stationary phase equations show that $y_{kl}=\epsilon_{kl} \frac{\xi_{kl}}{|\xi_{kl}|}$ and 

\begin{equation}
\begin{aligned}
\epsilon_{12}(2j_{12}+1)\delta_{12}+\epsilon_{13}(2j_{13}+1)\delta_{13}+\epsilon_{14}(2j_{14}+1)\delta_{14}+\epsilon_{15}(2j_{15}+1)\delta_{15}
& = & 0 \\
-\epsilon_{12}(2j_{12}+1)\delta_{12}+\epsilon_{23}(2j_{23}+1)\delta_{23}+\epsilon_{24}(2j_{24}+1)\delta_{24}+\epsilon_{25}(2j_{25}+1)\delta_{25}
& = & 0 \\
-\epsilon_{13}(2j_{13}+1)\delta_{13}-\epsilon_{23}(2j_{23}+1)\delta_{23}+\epsilon_{34}(2j_{34}+1)\delta_{34}+\epsilon_{35}(2j_{35}+1)\delta_{35}
& = & 0 \\
-\epsilon_{14}(2j_{14}+1)\delta_{14}-\epsilon_{24}(2j_{24}+1)\delta_{24}-\epsilon_{34}(2j_{34}+1)\delta_{34}+\epsilon_{45}(2j_{45}+1)\delta_{45}
& = & 0 \\
-\epsilon_{15}(2j_{15}+1)\delta_{15}-\epsilon_{25}(2j_{25}+1)\delta_{25}-\epsilon_{35}(2j_{35}+1)\delta_{35}-\epsilon_{45}(2j_{45}+1)\delta_{45}
& = & 0 \\
\end{aligned}
\end{equation}

These configurations will contribute to the integral whenever the above set of equations can be satified. This only happens for special values of $j$; generic 10j-symbols will not have stationary points of this type.

For example of when this type of stationary point does occur consider the regular 10j-symbol with all $j$'s equal. Arrange the $x$ vectors in a plane so that moving clockwise round the vectors they are in ascending order. Then all bivectors are equal and every $\delta$ is $+1$. 

Choosing $\epsilon_{13}=-1$, $\epsilon_{14}=-1$, $\epsilon_{25}=-1$, $\epsilon_{35}=-1$, and all other $\epsilon$'s equal to $+1$ gives one solution to the stationary phase equations.

We have four degrees of freedom to choose the angles from $x_1$ to the other four $x_i$ in the plane, and for all of these configurations the action will be stationary. The action will be zero for all of these configurations, so the contribution to the integral will be a constant multiplied by $\alpha^{-\frac{5}{2}}$.

\subsection{Limit of admissibility}

A vertex of a spin network is said to be at the limit of admissibility if the spins labelling the edges incident to it only just satisfy one of the admissibility inequalities for that vertex. For a vertex of a 4-simplex one of the spins is equal to the sum of the other three, for example
\begin{equation}\label{limit}
j_{12}=j_{13}+j_{14}+j_{15}.
\end{equation}

At the limit of admissibility the behaviour of the stationary point equations is typically different to the generic behaviour outlined above. However with the scaling given by $2j+1=\alpha(2J+1)$ for some constant $J$, this is not apparent because the scaling does not preserve the condition (\ref{limit}). Thus the only way to investigate the behaviour at the limit of admissibility in the asymptotic limit is to use a different scaling, namely
$$j=\alpha J$$
with constant $J$ and $\alpha\to\infty$.
The effect this has on the stationary phase formalism is to change the action (the part multiplying $\alpha$) to 
\begin{equation} \label{eq:actiont}
S=\sum_{k<l} (2j_{kl})\xi_{kl}\cdot y_{kl},
\end{equation} 
instead of (\ref{eq:actions}). Consequently the stationary point equations are the same but with $(2j+1)$ everywhere replaced with $2j$. The following considerations indicate how the above types of stationary point behave at the limit of admissibility with this alternative scaling.

\begin{figure}[h]
\begin{center}
\includegraphics{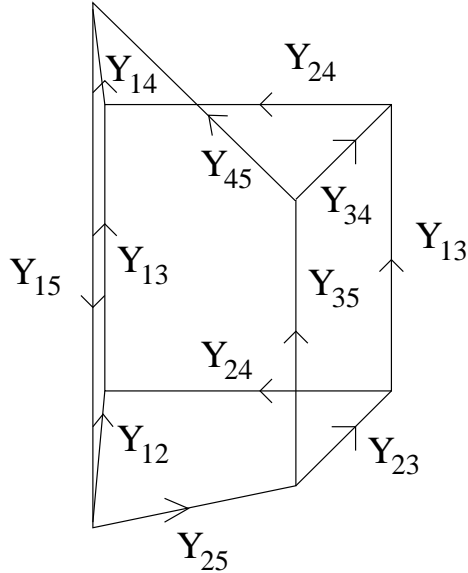}
\end{center}
\caption{4-simplex totally parallel stationary phase point - vertex 1 at limit of admissibility} \label{fig:pd}
\end{figure}

The asymptotics of the one-dimensional configurations is modified in this situation.
If one or more vertices of the 10j symbol are at the limit of admissibility the equation corresponding to that vertex will require the $y$'s to be parallel. If only one vertex is at the limit of admissibility (see figure \ref{fig:pd}) then the degrees of freedom available in deforming the cube will drop in the asymptotic limit from 5 to 2. The contribution in this case will be $\alpha^{-\frac{9-2}{2}}=\alpha^{-\frac{7}{2}}$, and may no longer be the leading contribution to the limit.

If more than one vertex is at the limit of admissibility there will in general be no solutions to the stationary phase equations with all $x$ parallel.

For the partly-parallel configuration with $x_1$ and $x_2$ parallel the equations (\ref{eq:pair}) show that
if vertex 1 is at the limit of admissibility then vertex 2 must also be.
   This stationary point is expected to give the leading-order contribution to the asymptotic formula of $\alpha^{-3}$ in the case where two vertices are at the limit of admissibility. This agrees with the numerical results in \cite{baez-chr:10j}.

If two pairs of $x$ vectors are separately parallel, the equations are compatible with two vertices at the limit of admissibility but the contribution decays as  $\alpha^{-\frac{7}{2}}$ and so this case does not give the leading contribution to the asymptotic formula.

 With four $x$ vectors parallel, the stationary equations (\ref{eq:fourp}) become
 \begin{equation} 
\epsilon_{15}j_{15}+\epsilon_{25}j_{25}+\epsilon_{35}j_{35}+\epsilon_{45}j_{45}=0
\end{equation}
 with this scaling. Some choices of $\epsilon$ reproduce the limit of admissibility condition for vertex 5. This gives a term of order $\alpha^{-3}$ in the case of one vertex at the limit of admissibility. This is the leading order contribution, again agreeing with the numerical evidence in \cite{baez-chr:10j}. The contribution will die off more quickly if more than one vertex is at the limit of admissibility.

%% file: chapter4.tex
\section{Lorentzian 10j Symbol}

\subsection{Evaluating SO(3,1) 10j-symbols}

In this section the relativistic spin networks for the Lorentz group $\SO(3,1)$ are analysed in a similar fashion to the Euclidean case. In the definition one replaces constructions related to $\SO(4)$ with the analogous concepts for $\SO(3,1)$. In particular, the spin labels for the Lorentz group are non-negative real numbers. The label for the $kl$-th edge is denoted $p_{kl}$.
There is an integral formula for $\SO(3,1)$ spin networks \cite{lsm} called the regularized evaluation, which takes the following form for the 10j-symbol (5 vertices and 10 edges). 
This integration is carried out over Hyperbolic space $H^3$ (the three dimensional subspace $x.x=-1$ of $-+++$ signature Minkowski space), instead of $S^3$ for the $\SO(4)$ case studied previously. The distance $r_{kl}$ between two variables $x_k, x_l\in H^3$ is defined by $\cosh r_{kl}=-x_k\cdot x_l$.
The evaluation is \begin{equation} \label{eq:li}
I=\frac{1}{(2 \pi^2)^4} \int_{x \in (H^3)^4} \prod_{k<l} \frac{\sin(p_{kl}r_{kl})}{\sinh(r_{kl})}dx_2dx_3dx_4dx_5
\end{equation}
Again there is a vector $x$ for each vertex of the spin network, and $x_2 \ldots x_5$ are integration variables. For this formula, $x_1$ is fixed at $(1,0,0,0)$, removing a symmetry which would otherwise multiply the integral by the infinite volume of hyperbolic space. Note that the formula in \cite{lsm} for the evaluation has been multiplied by  $\prod_{k<l} p_{kl}$ so that the analogy with the previous sections is clearer. 
Given two points $x_k$ and $x_l$ in $H^3$, it is possible to define a vector tangent to $x_k$ which indicates the position of $x_l$. Define $\xi_{kl}$ to be the initial velocity of the geodesic $\gamma(t)$ which has endpoints $\gamma(0)=x_k$ and $\gamma(1)=x_l$. Then $|\xi_{kl}|=r_{kl}$, the distance between $x_k$ and $x_l$. With this definition, $\xi_{kl}\in TH_{x_k}^3$. However it is possible to identify all of these tangent spaces if desired using an appropriate trivialisation of the tangent bundle.

Using the Kirillov formula 
$$\sin(p_{kl}r_{kl})=p_{kl}r_{kl} \int_{S^2} e^{ip_{kl} \xi_{kl} .y_{kl}}\frac{dy_{kl}}{4 \pi}$$
 and replacing $p_{kl}$ by $\alpha p_{kl}$ to effect the scaling gives
\begin{multline}
I=\frac{1}{(2 \pi^2)^4 (4 \pi)^{10}}\\
 \int_{x \in (H^3)^4} \int_{y \in \left(S^2\right)^{10}} \prod_{k<l} \frac{\alpha_{kl} p_{kl} r_{kl}}{\sinh(r_{kl})} e^{i \alpha \sum_{k<l} p_{kl}(\xi_{kl}. y_{kl})} dy dx_2dx_3dx_4dx_5.
\end{multline}

We use the stationary phase method as before to examine the asymptotics as $\alpha \to \infty$. It is necessary to find the stationary points of the action $S=\sum_{k<l} p_{kl}(\xi_{kl}. y_{kl})$ with respect to variations in $x$ and $y$.

\subsection{Results}
The stationary points can be classified in the same way as for the $\SO(4)$ case. The stationary phase equations for $\SO(3,1)$ are more or less exactly the same as for $\SO(4)$, except the vectors are in Minkowski space rather than four-dimensional Euclidean space. One qualitative difference is that the $x_i$ are future-pointing unit timelike vectors and so the symmetry $x_i\mapsto -x_i$ is absent. Nevertheless there exist non-degenerate, 4-dimensional, stationary points as demonstrated in detail below. These correspond to 4-simplexes in Minkowski space in which all tetrahedral faces are spacelike. A new feature is that some of these faces are future-pointing and some are past pointing. However the scaling behaviour of these stationary points is exactly the same as for the 
$\SO(4)$ case. Similary the other generic cases of stationary points exist in the same way and exhibit exactly the same scaling as for $\SO(4)$. In particular, the one-dimensional stationary points again scale as $\alpha^{-2}$ and again dominate the generic configuration in the asymptotic limit.

A new feature in the case of non-generic spins is the possibility of stationary points lying on the boundary of hyperbolic space. These will correspond to the possibility of tetrahedral faces of the geometric 4-simplex which are null. Clearly null tetrahedra cannot occur in the $\SO(4)$ case, where vanishing 3-volume implies that a tetrahedron is degenerate. However we do not analyse this interesting possibility in any further detail here.

For most of the types of stationary point the equations are the same as for the $\SO(4)$ case, and so they are not repeated here. The one case which is analysed in detail is the non-degenerate case in the next section.

\subsection{Non-degenerate 4-dimensional configurations}

Consider configurations where the $x_i$ are non-parallel and span 4 dimensions.

Varying the action with respect to the y's we find that the action is stationary when $y_{kl}$ is parallel or antiparallel to $\xi_{kl}$. Let $\epsilon_{kl}=\pm1$ then
\begin{equation}
y_{kl}=\epsilon_{kl} \frac{\xi_{kl}}{|\xi_{kl}|}
\end{equation}
The action at the stationary points in the $y$'s becomes
\begin{equation}
S=\sum_{k<l}\epsilon_{kl} p_{kl} r_{kl}
\end{equation}
To examine the contribution from non-degenerate stationary points we need a Schl\"afli identity for Lorentzian 4-simplexes in which every 3-dimensional face is spacelike. The normal vectors to these faces are therefore timelike and we need a definition of the angle between them.

\begin{description}
\item[Definition: Lorentzian angle \cite{barrettfoxon}] The Lorentzian angle bet\-ween two time\-like unit vectors $n_k$ and $n_l$, denoted by $\Theta_{kl}$, is defined in two distinct cases
\begin{enumerate}
\item Interior Lorentzian angle: If one of $n_k$ and $n_l$ is future pointing and the is other past pointing then $\Theta_{kl} \geq 0$ and is given by
\begin{equation}
\Theta_{kl}=\cosh^{-1}(n_k.n_l)
\end{equation}

\item Exterior Lorentzian angle: If $n_k$ and $n_l$ are both future or past pointing then $\Theta_{kl} \leq 0$ and is given by
\begin{equation}
\Theta_{kl}=-\cosh^{-1}(-n_k.n_l)
\end{equation}
\end{enumerate}

Setting $m_{kl}=0$ for exterior Lorentzian angles and $m_{kl}=1$ for interior Lorentzian angles we can combine these definitions into
\begin{equation} \label{eq:lorang}
\Theta_{kl}=-(-1)^{m_{kl}}\cosh^{-1}(-(-1)^{m_{kl}}n_k.n_l).
\end{equation}
This implies
$$n_k.n_l=-(-1)^{m_{kl}}\cosh(\Theta_{kl})$$
for all $k,l$, as long as $\Theta_{kk}$ is defined to be 0.
\end{description}

\newtheorem{theorem}{Theorem}

\begin{theorem}[Lorentzian Schl\"afli identity]
For a Lorentzian 4-simplex with areas $A_{kl}$ and timelike normal vectors $n_1 \ldots n_5$ 
\begin{equation}
\sum_{k<l}A_{kl} \, \dd\Theta_{kl}=0
\end{equation}
\end{theorem}

\begin{description}
\item[Proof:] This follows the derivation of Schl\"afli identity in \cite{barrett:regge} but using Lor\-entz\-ian instead of Euclidean geometry.
Let $\sigma$ be a 4-simplex with timelike normal vectors $n_1 \ldots n_5$ and Lorentzian 4-volume $|\sigma|$, $\sigma_i^3$ be the tetrahedron opposite vertex $i$ with 3-volume $T_i$ and $\sigma_{ij}^2$  the triangle of $\sigma$ opposite vertices $i$ and $j$ with area $A_{ij}$.

Stokes' theorem gives 
\begin{equation*}
\sum_{k}T_k n_k=0 
\end{equation*}
and dotting this with $n_l$ implies
\begin{equation*}\sum_k -(-1)^{m_{kl}}\cosh(\Theta_{kl})T_k=0  
\end{equation*}

Differentiating, and contracting with the vector $T_{l}$ gives
\begin{equation}\label{eq:ls1}
\sum_{k\neq l}T_k T_l\sinh|\Theta_{kl}| \, \dd \Theta_{kl}=0
\end{equation}

However, one can show that

\begin{equation}
T_k T_l \sinh|\Theta_{kl}|=\frac{4}{3} A_{kl} |\sigma|.
\end{equation}

This follows from the equations
$$ |\sigma|=\frac14 T_k h$$
$$ T_l=\frac13 h' A_{kl}$$
where $h$ and $h'$ are the altitudes of the 4-simplex and tetrahedron $T_l$ respectively. These are related by $h=h'\sinh|\Theta_{kl}|$.

Now equation (\ref{eq:ls1}) simplifies to

\begin{equation}
\sum_{k<l}A_{kl} \, \dd \Theta_{kl}=0. \end{equation}
\begin{flushright}$\Box$
\end{flushright}

\end{description}

Consider a geometric 4-simplex with timelike normals $n_1 \ldots n_5$. Rotate the 4-simplex so that $n_1$ is at $(1,0,0,0)$, $n_2$ is in the plane spanned by $(1,0,0,0)$,$(0,1,0,0)$ and $n_3$ is in the hyperplane spanned by $(1,0,0,0)$, $(0,1,0,0)$, $(0,0,1,0)$.

Now some of the normal vectors will be future pointing, and others will be past pointing. However the definition of the $x_{i}$ vectors in equation (\ref{eq:li}) requires them to be future pointing. Take $x_i=a_i n_i$ where $a_i=1$ if $n_i$ is future pointing and $a_i=-1$ if $n_i$ is past pointing. From the choice of alignment of the 4-simplex we always have $a_1=1$. These definitions give
\begin{equation}
\dd r_{kl}=a_k a_l \, \dd \Theta_{kl}
\end{equation}

Varying the action, and using a Lagrange multiplier $\lambda$ for the constraint

\begin{equation}
\dd S=\sum_{k<l}\epsilon_{kl} p_{kl}a_k a_l  \, \dd \Theta_{kl}=\lambda \sum_{kl}A_{kl} \, \dd \Theta_{kl}
\end{equation}
Set $\lambda=\pm1$ to fix the scale of the 4 simplex. Then we can identify
\begin{equation}
p_{kl}=A_{kl}
\end{equation}
and
\begin{equation}
\epsilon_{kl} = \lambda a_k a_l
\end{equation}
which fixes the $\epsilon$'s up to an overall sign.

The expected contribution from these configurations is an oscillating function multiplied by $\alpha^{-\frac{9}{2}}$. The stationary phase calculations are similar to the SO(4) case but now the sign of $a_k$ is determined by whether face $k$ is future or past pointing. Not all $a_k=1$ or $-1$ is possible since the 4 simplex must have at least one face future pointing and past pointing.

%% file: conclusion.tex
\section{Conclusions}

The stationary phase technique has been used to calculate asymptotic formulae for $\SO(4)$ relativistic spin networks.

The asymptotic formula for the relativistic spin network tetrahedron agrees exactly with the square of the Ponzano-Regge formula whenever we have a tetrahedron which can be embedded into 3-dimensional Euclidean space.

For the $\SO(4)$ relativistic spin network 4-simplex there are a number of types of stationary points in the asymptotic formula. For a generic set of spins the evaluation of the 10j-symbol is dominated in the asymptotic limit by degenerate one-dimensional geometries, whose contribution in general decays as a constant times $\alpha^{-2}$. 

The stationary points for the $\SO(3,1)$ relativistic spin network 4-simplex can be classified in the same way as for the $\SO(4)$ case. There exist non-degenerate, 4-dimensional, stationary points which correspond to 4-simplexes in Minkowski space in which all tetrahedral faces are spacelike. The stationary points corresponding to degenerate one-dimensional geometries again scale as $\alpha^{-2}$ and dominate the generic configuration in the asymptotic limit.